\documentclass[amsmath,pra,aps,noeprint,twocolumn,titlepage,nofootinbib,
preprintnumbers,floatfix,10pt]{revtex4-1}%

\usepackage{amsmath}
\usepackage{amsfonts}
\usepackage{amssymb}
\usepackage{graphicx}%
\usepackage{color}

\newcounter{RSQ}

\begin{document}

\title{Virtual Delbr\"uck scattering and the Lamb shift in light hydrogen-like atoms}

\author{Robert Szafron}
\email{robert.szafron@tum.de}
\affiliation{Technische Universit\"at M\"unchen, Physik-Department T31, 85748 Garching, Germany}

\author{Evgeny Yu. Korzinin}
\author{Valery A. Shelyuto}
\affiliation{D.~I. Mendeleev Institute for Metrology, St.Petersburg,
190005, Russia}
\affiliation{Pulkovo Observatory, St.Petersburg, 196140, Russia}
\author{Vladimir G. Ivanov} 
\affiliation{Pulkovo
Observatory, St.Petersburg, 196140, Russia}

\author{Savely~G.~Karshenboim}
\email{savely.karshenboim@mpq.mpg.de}
\affiliation{Ludwig-Maximilians-Universit{\"a}t, Fakult{\"a}t f\"ur Physik, 80799 M\"unchen, Germany}
\affiliation{Max-Planck-Institut f\"ur Quantenoptik, Garching, 85748, Germany}
\affiliation{Pulkovo Observatory, St.Petersburg, 196140, Russia}


\preprint{TUM-HEP-1209/19}

\begin{abstract}
We return to the problem of evaluation of the light-by-light contribution to the energy levels of the hydrogen atom. We find an additional contribution directly related to the Delbr\"uck scattering amplitude. The new correction is larger than the previously included light-by-light terms at order~$\alpha^2 (Z\alpha)^6\ln(Z\alpha)\, m_e$. We consider the effective potential in position space using an effective field theory approach and evaluate light-by-light corrections to the energy levels of states with non-zero orbital momentum as well as to the weighted difference of $s$ states. We also determine the large distance asymptotic behaviour of the effective potential induced by the light-by-light scattering in muonic atoms.   
\end{abstract}
\maketitle

\section{Introduction\label{s:introduction}}

Energy levels of hydrogen are one of the best-studied observables. Experiments can measure transition frequencies between different energy levels with astonishing accuracy \cite{Parthey:2011lfa,Matveev:2013orb} placing them at the frontier of precision physics. Theory of the hydrogen spectrum is also full of beautiful and non-trivial results. This increasing precision, as well as the renowned proton radius puzzle \cite{Pohl:2013yb,Higinbotham:2015rja,Carlson:2015jba}, demands further scrutiny of the hydrogen spectrum; and in particular, the higher-order radiative corrections must be carefully investigated and systematically evaluated.

Radiative corrections to the energy level $\Delta E(ns)$ can be parametrized as double expansion in the fine structure constant $\alpha$ and velocity of the electron $v\sim Z\alpha$, where $Z$ is the atomic number of the nucleus. The coefficients of expansion in $\alpha$ are defined through
\begin{eqnarray}\label{eq:F123}\nonumber
&&\Delta E(ns)=\frac{\alpha(Z\alpha)^4m_e}{\pi n^3}\times\\
&&\left(F^{(1)}+\frac{\alpha}{\pi}F^{(2)}+\left(\frac{\alpha}{\pi}\right)^2F^{(3)}+\ldots\right)\,.
\end{eqnarray}
We neglect here recoil corrections, i.e. we take the limit of infinite nucleus mass, $m_N\to\infty$. In this approximation, the nucleus is regarded as a point-like source of a static Coulomb potential.
For hydrogen and other light atoms, the coefficients $F^{(k)}$ can be further expanded in powers of~$Z\alpha$. In this work, we shall focus on the second-order corrections that are conventionally parameterized as
\begin{widetext}

\begin{equation}\label{eq:F2}
 F^{(2)}(Z\alpha)=B_{40}+\left(Z\alpha\right) B_{50}+\left(Z\alpha\right)^{2}\left(B_{63}\ln^3\left[(Z\alpha)^{-2}\right]+ B_{62}\ln^{2}\left[\left(Z\alpha\right)^{-2}\right]+B_{61}\ln\left[\left(Z\alpha\right)^{-2}\right]+B_{60}\right)+\ldots,
\end{equation}
\end{widetext}
where the $B_{ab}$ coefficients depend on the state of hydrogen. 
For a review of different terms in the expansion and theory of hydrogen spectrum see \cite{Eides:2000xc, eides2007theory,Mohr:2012tt,Yerokhin:2018gna}. Values of many known parameters can be found in \cite{Mohr:2015ccw} and in recently published works \cite{Karshenboim:2018mtf, Czarnecki:2016lzl}.

In a newly published letter \cite{Karshenboim:2019iuq}, new results became available for coefficients\footnote{Coefficients $C_{ab}$ are defined analogously to $B_{ab}$ coefficients, i.e., $F^{(3)}(Z\alpha)=C_{40}+\left(Z\alpha\right) C_{50}+\ldots$ }  $B_{60}$, $B_{61}$ and $C_{50}$, which reduced the overall uncertainty of the two- and three loop non-recoil corrections to the 1$s$ Lamb shift by a factor of three. Here we provide details of our computation for the coefficient $B_{61}$ and include additional results relevant for states with higher angular momentum and also for muonic atoms.

Two-loop quantum electrodynamics (QED) effects can be divided into different classes of diagrams. The pure self-energy diagrams constitute one such group. For these, in addition to analytical results \cite{Jentschura:2005xu,Pachucki:2001zz,Pachucki:2003gt,Czarnecki:2005sz,Dowling:2009md}, there exist numerical results for the self-energy corrections evaluated for $Z \geq 10$ \cite{Yerokhin:2003pq,yerokhin2003evaluation,yerokhin2009two}. A class of diagrams with closed electron loops is also known numerically \cite{yerokhin2008two}, but only in the free-loop approximation, i.e., when the external Coulomb field is not included in the propagator of the electron inside the loop. In particular, this means that the light-by-light diagrams are not included in the numerical study.

In this paper, we concentrate on the contribution due to the Compton scattering on the nucleus, also known as Delbr\"uck scattering \cite{Milstein:1994zz},  that is a part of the light-by-light (LbL) correction. Previously, the effect of Delbr\"uck scattering was investigated for the bound electron $g$-factor \cite{Karshenboim:2002jc,Czarnecki:2017kva}, but its contribution to the Lamb shift has not been considered thus far. LbL scattering is a non-linear effect in quantum electrodynamics, and it arises from diagrams with a closed electron loop with four photon attachments.

Our main result is a new contribution to the logarithmic coefficient $B_{61}$ in (\ref{eq:F2}), which was omitted in \cite{Pachucki:2001zz, Jentschura:2005xu, Czarnecki:2016lzl}.
We also verify previous, partial result \cite{Czarnecki:2016lzl} (see also \cite{Szafron:2017guu}) on the LbL radiative correction to the energy spectrum and extend it to states with non-zero orbital momentum.  Evaluation of various loop integrals that appear in our computations was performed with the help of computer programs: FIRE \cite{Smirnov:2014hma}, FIESTA \cite{Smirnov:2015mct}, and PackageX \cite{Patel:2016fam}. 

To begin with, we briefly discuss the framework of our computations. Secondly, we consider the effective potential in position space that is subsequently used to evaluate corrections to the energy levels of hydrogen-like ions. Finally, we compute the large distance asymptotic behaviour of the effective potential in muonic atoms. Throughout this paper, we use natural units $\hbar =c =1$ and the fine structure constant $\alpha = \frac{e^2}{4\pi}$.

\section{Effective field theory approach}
To evaluate the contribution of LbL scattering diagrams, we use an effective field theory (EFT) framework. This method exploits scale separation to disentangle long- and short-distance corrections. High-energy modes are integrated-out, and the short-distance physics is encoded in the matching coefficients of the low-energy operators. The non-relativistic EFT approach also allows for a systematic expansion in a small parameter, the velocity of the electron in the hydrogen-like ion.

Studies of non-relativistic bound states require a two-step approach, owing to the presence of two low-energy scales, electron momentum $\sim  m_e \, v$ and energy $\sim  m_e \, v^2$. First, QED is matched on the low-energy EFT known as non-relativistic QED (NRQED) \cite{Kinoshita:1990ai} (see also \cite{Labelle:1996uc,Labelle:1996en,Paz:2015uga}), which contains both soft, potential and ultra-soft modes. Full NRQED Lagrangian up to terms $1/m_e^4$ can be found in \cite{Hill:2012rh}; here we shall only give operators relevant to the LbL diagrams. To achieve homogeneous power counting in electron velocity, the soft and potential photon modes must be integrated out together with the soft electron modes and the resulting theory is known as potential non-relativistic QED (PNRQED) \cite{Pineda:1997ie, Pineda:1998kn,Beneke:1999zr} (for an alternative procedure, see \cite{Manohar:2000rz} or  \cite{Pachucki:2004zz}). PNRQED contains potential electron modes and ultra-soft photons. 

\begin{figure}[ht]
\includegraphics[width=0.5\columnwidth]{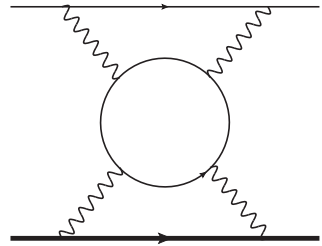}
\caption{\label{fig:LBL} The light-by-light  scattering contribution to the hydrogen energy spectrum. The thick line represents the nucleus in the zero-recoil approximation; the thin line is the electron. In this paper, we target the case where the upper loop is soft, while the remaining two loops are hard. }
\end{figure}

The light-by-light contribution to the hydrogen spectrum was previously considered in \cite{Eides:1994rc,Eides:1994dq,Pachucki:1993zz,Pachucki:1994ega, Dowling:2009md}. In that case, all the loop momenta are hard ($k \sim m_e$) and the diagram shown in Fig. \ref{fig:LBL} contributes to the matching coefficient of a local four-fermion operator
\begin{equation}\label{eq:d2}
 \mathcal{L}_{\textrm{ NRQED}}\supset d_2\, \psi_{e}^{\dagger}\psi_{e} N^\dagger N\,,
\end{equation}
where $\psi_{e}$ ($N$) is non-relativistic electron (nucleus) field and $d_2$ is a Wilson coefficient; symbol $\supset$ is used here to denote that an operator on the right side is a part of the full NRQED Lagrangian.  This operator is suppressed by $m^{-2}_e$, and the induced correction to the Dirac energy levels contributes to the $B_{50}$ term.

\begin{figure}[ht]
\includegraphics[width=0.8\columnwidth]{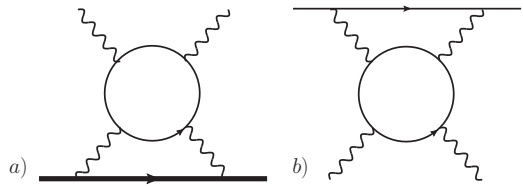}
\caption{\label{fig:LBL_C} Examples of two-loop LbL QED diagram contributions to the Compton amplitude necessary to obtain matching coefficients of operators in eq.~(\ref{eq:op_nr}) (left diagram) and eq.~(\ref{eq:op_LBL_1}) (right diagram).}
\end{figure}

The case when the momentum in the upper loop is hard, while the lower one is soft $k\sim  m_e \, v $,  was considered before in~\cite{Czarnecki:2016lzl}. In this instance, the diagram on the right in Fig. \ref{fig:LBL_C} contributes to the matching coefficients of operators\footnote{Note that this form is a direct consequence of QED gauge invariance. We may rewrite  eq.~(\ref{eq:op_LBL_1}) as
$$ 
\mathcal{L}_{\textrm{ NRQED}}\supset\psi_{e}^{\dagger}\left[\frac{C'_{A1}}{2} F^{\mu \nu}F_{\mu \nu}+C'_{A2}u_\mu u^\nu F^{\mu \sigma}F_{\nu \sigma} \right]\psi_{e}\, , $$
where four-vector $u$ is defined below eq.~(\ref{eq:Delbruck}). In this form, the gauge invariance is manifest.  }
\begin{equation}\label{eq:op_LBL_1}
\mathcal{L}_{\textrm{ NRQED}}\supset\psi_{e}^{\dagger}\left[C'_{A1}\left({\mathbf B}^{2}-{\mathbf E}^{2}\right)-C'_{A2}{\mathbf E}^{2}\right]\psi_{e}\,,
\end{equation}
where ${\mathbf E}$ (${\mathbf B}$) is the electric (magnetic) field that contains soft, potential and ultra-soft modes. $C'_{A1}$ and $C'_{A2}$ are Wilson coefficients whose sum was evaluated in \cite{Czarnecki:2016lzl}. The above operator is suppressed by $m_e^{-3}$. Different mass scaling explains why this correction modifies coefficient $B_{61}$ as opposed to the pure hard loop correction that starts one order lower in the $Z\alpha$ expansion. 

In this paper, we concentrate on the case when the upper loop is soft, and the lower ones are hard. Naively, one could expect that in this case the Wilson coefficients of operators
\begin{equation}\label{eq:op_nr}
 \mathcal{L}_{\textrm{ NRQED}}\supset N^{\dagger}\left[C_{A1}\left({\mathbf B}^{2}-{\mathbf E}^{2}\right)-C_{A2}{\mathbf E}^{2}\right]N\,,
\end{equation}
which are bilinear in the nucleus field, are suppressed by inverse powers of the nucleus mass. This, however, is not the case, as the leading contribution is obtained from the region of the loop momenta of the order of electron mass. The non-recoil limit used for the nucleus makes the diagrams in Fig.~\ref{fig:LBL_C} asymmetric, forcing the loop integral in the Delbr\"uck amplitude to be effectively three-dimensional, while the loop integral encapsulating the upper electron line in diagram b) is still four-dimensional with a full relativistic electron propagator.

The matching coefficients $C_{A1}$ and $C_{A2}$ can easily be inferred from the on-shell Delbr\"uck scattering amplitude in the low energy limit. Denoting the momenta of external photons as $k_1$ and $k_2$, we find the amplitude \footnote{A similar result was obtained before in Ref.~\cite{Karshenboim:2002jc}, where a different convention was used (in particular, the aforementioned reference uses $\alpha = e^2$). We note that the final result of the computation expressed entirely in terms of $\alpha$ does not depend on the relation between $\alpha$ and $e$.}
(right diagram in Fig. \ref{fig:LBL_C}) in the limit $k_1\sim k_2 \ll m_e \ll m_N$ (see also~\cite{Costantini:1971cj,Papatzacos:1975bt,Papatzacos:1975bu,Karshenboim:2002jc})
\begin{eqnarray}\label{eq:Delbruck}
 T^{\mu\nu}\left(k_{1},k_{2}\right)=e^{2}\frac{\left(Z\alpha\right)^{2}}{m_e^{3}}\left[C_{1}\left(g^{\mu\nu}k_{1}k_{2}-k_{2}^{\mu}k_{1}^{\nu}\right)+\right.  \nonumber\\ \left. C_{2}\left(w^{2}g^{\mu\nu}-w\left(u^{\mu}k_{1}^{\nu}+k_{2}^{\mu}u^{\nu}\right)+k_{1}k_{2}u^{\mu}u^{\nu}\right)\right]\, ,
\end{eqnarray}
where we introduced $u^\mu=\left(1,0,0,0\right)$ and $w$ is the photon energy $w=u\cdot k_1 = u\cdot k_2$. We assume that the $k_1$ momentum is incoming and the $k_2$ momentum is outgoing. 
The coefficients $C_i$ are
\begin{eqnarray}
 C_1 &=& \phantom{-}\frac{7}{1152},  \nonumber\\
 C_2 &=& - \frac{73}{2304},
\end{eqnarray}
and we have the following relation 
\begin{equation}
 C_{Ai}=e^2\frac{\left(Z\alpha\right)^{2}}{2m_e^{3}}C_{i},\;\;\; i=1,2\,. 
\end{equation}

\begin{figure}[ht]
\includegraphics[width=0.8\columnwidth]{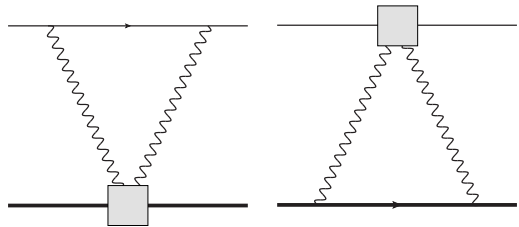}
\caption{\label{fig:LBL_V} One-loop matching diagrams of operators in eq.~(\ref{eq:op_nr}) (left diagram) and eq.~(\ref{eq:op_LBL_1}) (right diagram) represented by grey squares, on the effective potential in eq.~(\ref{eq:pot}) defined in PNRQED.}
\end{figure}

Having obtained the relevant matching coefficients of the NRQED Lagrangian, we perform the second matching step and integrate out the soft and potential photon modes. The resulting  effective PNRQED interaction Lagrangian $\mathcal{L}_{\text{\tiny POT}}$ is  a non-local four-fermion operator whose Wilson coefficient $V_{\textrm{LbL} }\left(\mathbf{r}\right)$ depends on $\mathbf{r}$ as $\left|\mathbf{r}\right|^{-4}$ 
\begin{equation}\label{eq:pot}
 \mathcal{L}_{\text{\tiny POT}}(x)=-\int d^{3}r\left[\psi_{e}^{\dagger}\psi_{e}\right]\left(x+{\mathbf r}\right)V_{\textrm{LbL} }\left(\mathbf{r}\right)\left[N^{\dagger}N\right]\left(x\right).
\end{equation}

Evaluating the two diagrams in Fig.~\ref{fig:LBL_V} we find contributions to the matching coefficient of the potential interaction 
\begin{equation}
 V^{(i)}_{\text{LbL}}\left(r\right)={\mathcal{C}}^{(i)}_{\text{LbL}}\left(\frac{\alpha}{\pi}\right)^{2}\frac{\left(Z\alpha\right)^{2}}{4m_e^{3}r^{4}},
\end{equation}
with $r=\left| \mathbf{r}\right| $.
The left diagram gives 
\begin{equation}
{\mathcal C}^{(a)}_{\text{LbL}}= 2\pi^2\left( C_1+C_2\right)=-\frac{59 \pi^2}{1152}.
\end{equation}
The contribution from the second diagram, previously evaluated in~\cite{Czarnecki:2016lzl}, is
\begin{equation}
{\mathcal C}^{(b)}_{\text{LbL}}= \frac{43}{36}-\frac{133}{864}\pi^2.
\end{equation}
We find the total long-distance LbL potential, i.e., for typical atomic distances much larger than the Compton wavelength of the electron, $r\sim 1/( Z\alpha \, m_e) \gg 1/m_e$, to be
\begin{equation}
 V_{\text{LbL}}\left(r\right)={\mathcal C}_{\text{LbL}}\left(\frac{\alpha}{\pi}\right)^{2}\frac{\left(Z\alpha\right)^{2}}{4 m_e^{3}r^{4}}\, ,
\end{equation}
with 
\begin{equation}
 {\mathcal C}_{\text{LbL}} = {\mathcal C}^{(a)}_{\text{LbL}}+{\mathcal C}^{(b)}_{\text{LbL}} = \frac{43}{36}-\frac{709 \pi ^2}{3456}\, .
\end{equation}
The new contribution ${\mathcal C}^{(a)}_{\text{LbL}}$ is more than 1.5 times larger than the previously evaluated ${\mathcal C}^{(b)}_{\text{LbL}}$.

\section{Corrections to the energy levels}

Having computed the potential, we evaluate corrections to the energy levels using standard quantum-mechanical perturbation theory. The matrix element of the ${r^{-4}}$ potential in the $ns$ state is logarithmically enhanced and therefore it contributes to the $B_{61}$ coefficient. We find 
\begin{equation}
 \left\langle \frac{1}{r^{4}}\right\rangle _{ns}=4\frac{\left(Z\alpha\right)^{4}}{n^3}\, m_e^{4}\ln(Z\alpha)^{2}+\ldots\,,
\end{equation}
where dots indicate terms without logarithmic enhancement and  $\left\langle \;\right\rangle _{ns}$ denotes a dimensionally regularized matrix element in the $ns$ state. 
This means that the coefficient $\mathcal{C}_{\rm LbL}$ is related to the $B_{61}$\, as follows:
\begin{equation}
  B^{\text{LbL}}_{61}(ns) = - \mathcal{C}_{\text{LbL}}\,.
\end{equation}
Besides, there is a divergent contribution\footnote{The powerlike divergent integrals vanish in dimensional regularization. They are related to a short-distance $\delta$ potential which contributes to the $B_{50}$ term, see eq.~(\ref{eq:d2}). Here we discuss only the  logarithmically divergent integral associated with the $B_{60}$ part.} related to the $B_{60}$ term, see e.g. \cite{Jentschura:2005xu,Czarnecki:2016lzl}. As was discussed in~\cite{Czarnecki:2016lzl}, this divergence cancels after diagrams with one additional photon connecting the electron and nucleus lines are included. 
The total logarithmic $B_{61}$ LbL correction decreases the $1s-2s$ energy split by 720~Hz, out of which 440~Hz is due to Delbr\"uck scattering. 

Interestingly, the sign of the LbL contribution to the~$B_{61}$ term is opposite to the contribution of the corresponding diagram to the $B_{50}$ term. As an aside, we explain this fact by a simple, yet not a very formal argument based on the short-distance behaviour of the hydrogen wave function. Although we do not have a  proof of this fact, we expect that the LbL kernel is a well-behaved function whose sign is determined by the first term in the low-momentum expansion.  In coordinate space, the corrections are proportional to 
\begin{equation}\label{eq:R}
 \int d^3 r \frac{\left| R_{ns}\left(r\right) \right|^2}{r^4}\, ,
\end{equation}
where $R_{nl}\left(r\right)$ is the radial part of the non-relativistic hydrogen wave function in the $nl$ state. For $s$ states, the expansion of the wave function around the origin is 
\begin{equation}\label{eq:Rexp}
 R_{ns}\left(r\right) = R_{n0}\left(0\right)\left(1 - Z\alpha m_e r +\ldots \right)\, .
\end{equation}
The first term in the expansion is power divergent when inserted into (\ref{eq:R}) and gives rise to the $B_{50}$ term. The second term is $n$ independent and logarithmically divergent. It produces the $B_{61}$ correction with the opposite sign to the $B_{50}$ term. Both logarithmic corrections due to $V^{(a)}_{\text{LbL}}$ and $V^{(b)}_{\text{LbL}}$ are comparable in their magnitude. Moreover, the size of the logarithmic LbL correction can be estimated, following eq.~(\ref{eq:Rexp}), as  $$B^{(a)}_{61}\sim B^{(b)}_{61}\sim\ -B_{50}\, m_e \, r_0$$ for some $r_0$ of the order of the Compton wavelength of the electron. This argument also explains why both contributions (a) and (b) are similar in size. 

In~\cite{karshenboim1994jetp,Karshenboim:1996bg} it was observed that all hard contributions to the matrix element (in our case: the divergence related to $B_{50}$ and logarithmically divergent part of $B_{60}$) cancel in a specific difference of any two $ns$ state energy levels weighted by $n^3$,
\begin{equation}
 \Delta(n) = E({1s})-n^3 \, E(ns).
\end{equation}
Thus we also find the LbL correction to the difference of $ B_{60}\left(ns\right)$ and $B_{60}\left(1s\right)$ coefficients
\begin{eqnarray}\label{eq:diff}
 B_{60}\left(1s\right)-B_{60}\left(ns\right)=\nonumber\\
 -2\mathcal{C}_{\text{LbL}}\left[H_{n}-\ln n-\frac{2}{3}-\frac{1}{2n}+\frac{1}{6n^{2}}\right]\, ,
\end{eqnarray}
where $H_n=\sum_{k=1}^{n} \frac{1}{k}$ are harmonic numbers. Evaluation of the weighted difference $\Delta(n)$ requires a regularization of the integral appearing in the computation of the expectation value. This can be achieved, for example, by changing the dimensionality of the space. Once the results for $B_{60}\left(1s\right)$ and $B_{60}\left(ns\right)$ are combined, the regulator dependence cancels, and the result is finite.  

In Table \ref{tab:s}, we present numerical results for the difference in eq.~(\ref{eq:diff}) for several  precisely  measured hydrogen and deuterium ($Z=1$) states, relevant for the determination of the proton radius $r_p$. In the past, only the $2s-2p$  transition was measured precisely. Nowadays, thanks to the measurement of optical transitions frequencies of the main structure, many other transitions between energy levels can be measured \cite{Parthey:2011lfa, weitz1992m, weitz1994m,berkeland1995precise,bourzeix1996s, de1997absolute,schwob1999c,fleurbaey2018new,2018PhRvA}. Evaluation of the proton radius based on spectroscopic data relies on precise experimental data for several hydrogen and deuterium transitions \cite{Karshenboim:1997zu,deBeauvoir2000,Mohr:2000ie,Karshenboim:2014baa}. The spectroscopic measurements are compatible with the value of the proton radius determined from the elastic electron-proton scattering data  \cite{Arrington:2015ria}. However, the measurement of the proton radius based on muonic hydrogen \cite{antognini2013proton}, which has an essentially lowest uncertainty, shows a substantial discrepancy with electron based evaluation, thus motivating continuous progress in precise studies of the hydrogen spectrum both from the theoretical and experimental side. In particular, the computations of the Lamb shift demand further scrutiny.

\begin{table}
\caption{\label{tab:s} Contribution to the $B_{60}$ coefficients due to LbL correction. In the last column we present the LbL  contribution to the difference of the $B$ coefficients, $B_{60}(1s)-B_{60}(ns)$.}
 \begin{center}
\begin{tabular}{ccccc}
\hline 
\hline 
$n$  & $B_{60}\left(np\right)\,$ & $B_{60}\left(nd\right)$ & $B_{60}(1s)-B_{60}(ns)$\tabularnewline
\hline 

2 & $-$0.069193 & - & 0.11317\tabularnewline

3 &  $-$0.076881 & $-$0.0030752 & $-$0.13301\tabularnewline

4 &  $-$0.079571 & $-$0.0034596 & $-$0.13984\tabularnewline

6 &  $-$0.081493 & $-$0.0037342 & $-$0.14469\tabularnewline

8 &  $-$0.082166 & $-$0.0038303 & $-$0.14638\tabularnewline

12 &  $-$0.082647 & $-$0.0038989 & $-$0.14758\tabularnewline
\hline 
\hline 
\end{tabular}
\end{center}
\end{table}

Since we obtained the complete LbL part of the $B_{61}$ coefficient, we can also determine the LbL contribution to $B_{60}(nl)$, $l\neq0$. The same $1/r^4$ potential gives rise to the correction, but the non-relativistic wave function of states with $l\neq0$ vanishes at the origin and thus suppresses the UV behaviour leading to convergent matrix elements. The required integrals of the wave function multiplied by the $r^{-4}$ potential can be evaluated for arbitrary $n$ and $l$ \cite{landau1958course} using a standard representation of the hydrogen wave function with associated Laguerre polynomials.  The results do not depend on the total angular momentum $j$ and we find
\begin{eqnarray}\label{eq:B60p}
 B_{60}\left(nl\right)&=&\mathcal{C}_{\text{LbL}}\left(1s\right)\frac{n^{3}}{Z\alpha m_e}\int_{0}^{\infty}\frac{dr}{r^{2}}\frac{\left|R_{nl}\left(r\right)\right|^{2}}{\left|R_{1s}\left(0\right)\right|^{2}}\\
 &=& \mathcal{C}_{\text{LbL}} \frac{3 n^2-l^2-l}{l \left(8 l^4+20 l^3+10 l^2-5 l-3\right) n^2}\, . \nonumber
\end{eqnarray}
For example, for the $p$ ($l=1$) and $d$ ($l=2$) states we find
\begin{eqnarray}
B_{60}\left(np\right)	&=&\mathcal{C}_{\text{LbL}}\frac{3n^{2}-2}{30n^{2}} \,,\nonumber \\
B_{60}\left(nd\right)	&=&\mathcal{C}_{\text{LbL}}\frac{n^{2}-2}{210n^{2}}\, .
\end{eqnarray}
Numerical results for the corrections to the $B$ coefficients are shown in Table \ref{tab:s}.

Finally, we want to argue that our results (\ref{eq:diff}) and (\ref{eq:B60p}) constitute the complete LbL contribution up to higher-order corrections in $Z\alpha$. First, let us consider the case when only the closed electron loop is hard. In this case, we should match the LbL loop on the Euler-Heisenberg Lagrangian \cite{Heisenberg:1935qt}, which is suppressed by $m_e^{-4}$; thus, we expect this contribution to energy levels starts at the $\mathcal{O}\left(\alpha^2(Z\alpha)^7 m_e\right)$. Next, we should consider an ultra-soft contribution, i.e., a case when the photon momentum scales like $  m_e \, v^2$ and the electron propagator has to be replaced by the Coulomb Green's function. Note, however, that in PNRQED, the ultra-soft interaction $\psi_e^\dagger(t,{\mathbf x}) \, e \, {\mathbf x} \cdot \vec{E}(t,{\mathbf 0})\,\psi_e(t,{\mathbf x})  $ is velocity suppressed, and thus the ultra-soft contribution is also of the higher-order in $Z\alpha$. Indeed, the leading LbL correction is related to the $B_{50}$ coefficient, and it is convergent; therefore, any low-momentum input is suppressed by $Z\alpha$, as we see in the example of $B_{61}$ contributions. The hard loops present in our diagrams are also convergent, thus if we replace one of them with a soft or an ultra-soft contribution, the result will be further suppressed by $Z\alpha$.   

\begin{figure}[ht]
\includegraphics[width=0.5\columnwidth]{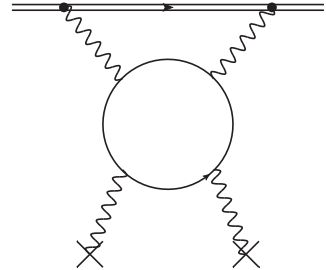}
\caption{\label{fig:fullQED} LbL loop evaluated with the electron in the external field. Unlike in the previous diagrams, here we use a double line to indicate that instead of  the free electron propagator, we use Coulomb Green's function.  }
\end{figure}

The fact that our result constitutes the full LbL correction is preeminently non-trivial for the $B_{60}(1s)-B_{60}(ns)$ difference. As a cross-check of our argument based on the power counting in PNRQED, we also examined the full NRQED diagram with the Coulomb electron propagator $G(\mathbf{r},\mathbf{r}',E)$  
\begin{equation}\label{eq:Coulomb}
G(\mathbf{r},\mathbf{r}',E) = \langle \mathbf{ r} | \frac{i}{E-H+i \epsilon}| \mathbf{r}' \rangle  = \sum_{n}i\frac{\psi_n(\mathbf{r})\psi_n^*(\mathbf{r}')}{E-E_n+ i\epsilon},
\end{equation}
where $\psi_n(\mathbf{r})$ are energy eigenstates in position representation, $H\psi_n(\mathbf{r}) = E_n \psi_n(\mathbf{r})$. The corresponding QED diagram is shown in Fig.~\ref{fig:fullQED}.
If we take $E\sim  Z\alpha\, m_e$,  we can neglect the non-relativistic Hamiltonian $H$ in the denominator of (\ref{eq:Coulomb}), and the resultant expression is equivalent to our computation with the free-electron propagator owing to the completeness of states relation. A similar argument was applied before in \cite{Karshenboim:2010cq,Karshenboim:2010cp} to evaluate the LbL contribution for the muonic atoms. 
On the other hand, if we assume that $E\sim (Z\alpha)^2 \, m_e$, then we need to include subleading terms in the energy expansion of the LbL loop, while still maintaining vanishing total energy transfer to the nucleus. These terms generate additional suppression, and so this region does not contribute to the $B_{60}$ term.

\section{Muonic hydrogen}
Our result is not directly applicable to muonic hydrogen. The reason is that the Bohr radius in light muonic atoms $1/r_\mu \sim  Z\alpha \, m_\mu$ is of the order of the Compton wavelength of the electron $r_\mu \sim 1/m_e$. Accordingly, the full dependence on the electron mass must be retained in the potential region for muonic hydrogen and other muonic atoms with low values of $Z$. We devoted a separate paper to this objective \cite{Korzinin:2018tnx}. Here we derive only the large distance asymptotic behaviour of the effective potential for the muonic atom, for which we can obtain an analytic result. 

\begin{figure}[t]
\includegraphics[width=0.95\columnwidth]{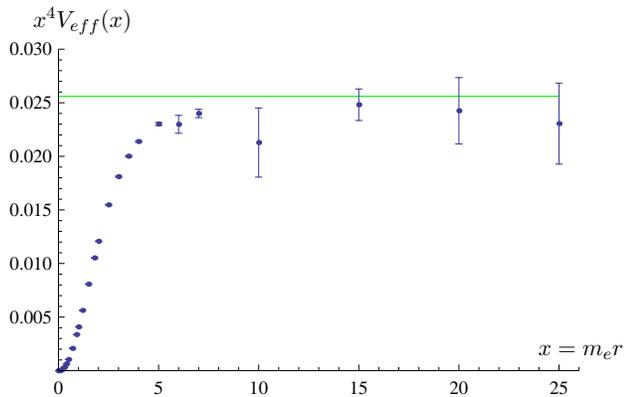}
\caption{\label{fig:plotMu} Potential for muonic atoms. We define dimensionless effective potential $V_{eff}(x)$ through  $-\alpha^2 (Z\alpha)^2 m_e V_{eff}(x) =V^{(\mu)}_{\text{LbL}}\left(r\right) $ and $x=m_e r$. Solid line is our analytic result (\ref{eq:PotMu}),  $x^4V_{eff}(x)\to  0.02561$.  Points are the result of numerical computation of the potential, keeping the full dependence on the electron mass. }
\end{figure}

As $r \to \infty$, it is permitted to apply a set-up similar to the one used in the hydrogen case. The hierarchy of relevant scales in this case is 
\begin{equation}
m_N \gg m_\mu \gg m_e \gg \frac{1}{r}\, .
\end{equation}
In consequence,  both nucleus and muon propagators should be expanded in the inverse powers of the heavy masses and thus the contribution related to the diagram (b) in Fig.~\ref{fig:LBL_C} is equal to the contribution from diagram (a) for muonic atoms in the large distance limit.  We find the effective LbL potential applicable to muonic atoms for $r\gg 1/m_e$
\begin{equation}\label{eq:PotMu}
V^{(\mu)}_{\text{LbL}}\left(r\right) \xrightarrow[r\to\infty]{} {2\mathcal{C}}^{(a)}_{\text{LbL}}\, \left(\frac{\alpha}{\pi}\right)^{2}\frac{\left(Z\alpha\right)^{2}}{4m_e^{3}r^{4}}\,.
\end{equation} 
We checked our asymptotic result by comparing it with the result of a direct numerical computation that keeps full dependence on the electron mass. The result is presented in Fig.~\ref{fig:plotMu}; we find a good agreement between these two methods. It is worth pointing out that our analytic computation is complementary to the numerical result, as the numerical accuracy decreases with $r$.

\section{Conclusions}
We evaluated the contribution to the Lamb shift in light hydrogen-like atoms due to the Delbr\"uck amplitude, and we found an additional contribution to the coefficient $B_{61}$. Including previously computed LbL corrections, the total LbL correction is
\begin{equation}
 B_{61}^{\text{LbL}}(ns) = 0.830\,309\, \ldots \,.
\end{equation}
The LbL logarithms were omitted before in \cite{yerokhin2008two} and \cite{Jentschura:2005xu}, and partially included in \cite{Yerokhin:2018gna} after publication of \cite{Czarnecki:2016lzl}.  Our result indicates that the uncertainty assigned to missing LbL contribution in \cite{yerokhin2008two} and also in \cite{Yerokhin:2018gna} was most likely underestimated. The detailed and most up-to-date discussion of  uncertainty of the overall $\alpha^8 m_e$ contribution can be found in \cite{Karshenboim:2019iuq}. 

We also corrected the weighted difference $\Delta(n)$ and evaluated the LbL corrections to states with $l\neq 0$ for general $n$ and $l$. In this instance, there is no logarithmic enhancement of the correction. We gave specific numerical results for the transitions that are most accurately measured. 
It is interesting to note that both LbL corrections have the same sign that is opposite to the LbL part of $B_{50}$. 

The LbL corrections were also analysed for the muonic atoms. In this case, we used the Delbr\"uck amplitude to determine the asymptotic behaviour of the potential in the large distance limit $r\to\infty$. The obtained result agrees with numerical evaluation and improves the accuracy of the potential at large values of $r$.

Our computations were performed in a modern framework based on non-relativistic EFT with potential interactions. This framework offers systematic power-counting and a clear separation of short and long-distance effects. We hope that this approach will become more popular in the future, allowing the cross-check of existing results and evaluation of higher-order corrections. 

 \acknowledgments

 The work was supported in part by DFG (Grant No. KA 4645/1-1). The
work on the effective potential for muonic atoms was supported in part by RSF
(under grant \# 17-12-01036). The authors are grateful to Andrzej Czarnecki, Aleksander Milstein, Akira Ozawa, Krzysztof
Pachucki, and Thomas Udem for useful and stimulating discussions.

\end{document}